\documentclass[12pt]{amsart}		

\usepackage{graphicx}        

\usepackage[all,cmtip]{xy}
\usepackage{amsmath, amsfonts, amsthm}
\usepackage{fullpage}

\newtheorem{thm}{Theorem}
\newtheorem*{thm*}{Theorem}

\theoremstyle{definition}

\theoremstyle{remark}
\newtheorem{rem}{Remark}
\newtheorem{eg}{Example}

\newcommand{\refeq}[1]{(\ref{#1})}

\newcommand{\beq}{\begin{equation}}
\newcommand{\eeq}{\end{equation}}
\newcommand{\bea}{\begin{eqnarray*}}
\newcommand{\eea}{\end{eqnarray*}}

\newcommand{\CC}{{\mathbb C}}
\newcommand{\Cset}{{\mathbb C}}
\newcommand{\Hset}{{\mathbb H}}
\newcommand{\Sset}{{\mathbb S}}
\newcommand{\RR}{{\mathbb R}}

\newcommand{\bb}{\mathbf{b}}
\newcommand{\bF}{\mathbf{F}}
\newcommand{\bg}{\mathbf{g}}
\newcommand{\bh}{\mathbf{h}}
\newcommand{\bu}{\mathbf{u}}
\newcommand{\bv}{\mathbf{v}}
\newcommand{\bx}{\mathbf{x}}
\newcommand{\bK}{\mathbf{K}}
\newcommand{\bm}{\mathbf{m}}

\newcommand{\fg}{\mathfrak{g}}

\newcommand{\cA}{{\mathcal A}}
\newcommand{\cB}{{\mathcal B}}

\newcommand{\cD}{{\mathcal D}}
\newcommand{\cE}{{\mathcal E}}
\newcommand{\cG}{{\mathcal G}}
\newcommand{\cM}{{\mathcal M}}
\newcommand{\cN}{{\mathcal N}}
\newcommand{\cR}{\mathcal{R}}

\newcommand{\Ga}{\Gamma}
\newcommand{\Om}{\Omega}
\newcommand{\om}{\omega}
\newcommand{\al}{\alpha}
\newcommand{\la}{\lambda}
\newcommand{\si}{\sigma}

\newcommand{\p}{\partial}
\newcommand{\half}{\frac{1}{2}}
\newcommand{\tr}{\mathrm{tr}\,}
\newcommand{\diag}{\mbox{diag}}

\begin{document}


\title{Dressing with Control:\\ Using integrability to generate\\ desired solutions to Einstein's equations}
%
\author{
Shabnam Beheshti}
\address{
Department of Mathematics\\
Rutgers, The State University of New Jersey\\
Piscataway, NJ, USA 08854
}
\email{beheshti@math.rutgers.edu}
\author{
Shadi Tahvildar-Zadeh
}
\address{
Department of Mathematics\\
Rutgers, The State University of New Jersey\\
Piscataway, NJ, USA 08854
}
\email{shadi@math.rutgers.edu}
\date{}
\begin{abstract}
Motivated by integrability of the sine-Gordon equation, we investigate a technique for constructing desired solutions to Einstein's equations by combining a  dressing technique with a control-theory approach.  After reviewing classical integrability, we recall two well-known Killing field reductions of Einstein's equations,  unify them using a harmonic map  formulation, and state two results on the integrability of the equations and solvability of the dressing system.  The resulting  algorithm is then combined with an asymptotic analysis to produce constraints on the degrees of freedom arising in the solution-generation mechanism.  The approach is carried out explicitly for the Einstein vacuum equations.  Applications of the technique to other geometric field theories are also discussed.
%
\end{abstract}

\maketitle



\section{Introduction}

Can solutions to the Einstein Equations be ``made to order," i.e., is it possible to construct spacetimes with a prescribed set of asymptotic ``observables" (such as total mass and angular momentum) and a prescribed causal and/or singular structure (e.g. number of components of event horizon, or number of ring singularities, etc.), given a specified set of initial parameters?  We provide a first step in answering this question by investigating a solution-generation mechanism  for integrable {\em harmonic maps}, known as the {\em vesture}, or {\em dressing} method.

%
%
%
%
%
The seeds of our discussion, however, are first sowed in sine-Gordon theory.  The special class of nonlinear equations to which the sine-Gordon equation belongs is that of classically integrable partial differential equations (PDEs).  Even though nonlinear, these PDEs have the property that new solutions can be formed from old ones, in what amounts to a nonlinear generalization of the superposition principle.  We exploit this very feature in order to study the Einstein equations.  As a consequence, the main thread of this communication can be summarized in the following diagram:
\[
\mbox{sine-Gordon} \longrightarrow \mbox{chiral fields} \longrightarrow \mbox{Einstein's equations} \longrightarrow \mbox{harmonic maps.} 
\]
We shall first briefly introduce classically integrable PDEs.  Then, building on the techniques developed for the sine-Gordon equation, we introduce Einstein's gravitational equations and put them in a similar context as sine-Gordon, by briefly  reviewing both the Zakharov-Belinski and Ernst formulations of the stationary axisymmetric Einstein vacuum equations.  We will then realize  this same system of PDEs as an axially symmetric harmonic map.

With background established, we place the gravitational field equations in the context of integrability by way of two theorems; informally, the results state that axially symmetric harmonic maps satisfy an integrable system of equations and that new axially symmetric harmonic maps with any number of prescribed singularities can be constructed from a given one.  We implement the algorithm afforded by the proofs of these theorems explicitly in the case of the Einstein vacuum and Einstein-Maxwell equations.
It is in the final section that we give a partial answer to the question posed at the beginning of this section, namely, by combining a novel asymptotic expansion with the calculated results of the vesture method, we show that, given any real numbers $M>0$ and $J$, it is indeed possible to treat the arbitrary constants of the dressing procedure as control parameters and produce a 1-solitonic harmonic map that, when viewed as a stationary axisymmetric spacetime metric, has total ADM mass equal to $M$ and total ADM angular momentum equal to $J$, and is otherwise free of unwanted pathologies (such as a non-zero NUT parameter), in other words, it is a member of the Kerr family of metrics.  We conclude by proposing a strategy for generating solutions in more general cases and indicate possible future directions of study.


\section{Integrability and Inverse Scattering}

A nonlinear PDE (or system of PDEs) is said to be {\em classically integrable} if there exists an overdetermined linear system, called a {\em Lax system},  the compatibility condition of which is precisely the nonlinear PDE\footnote{The term \emph{integrable} has been adopted in various contexts, and is ambiguous in characterizing the features of such an equation (e.g., existence of closed-form solutions by quadrature, possessing infinitely many conservation laws, exhibiting solitonic dynamics, etc.).  We shall restrict our attention to those equations which are integrable in the Lax sense.}.  We describe this in further detail.

Start with a nonlinear evolution equation for $u(t,x)$.  Associate to it a corresponding overdetermined system for  an isospectral family of linear differential operators $L(t,u)$, with $u$-dependent coefficients, whose eigenfunctions satisfy an additional linear evolution equation
\begin{equation}\label{eq:LB}
L\psi=\lambda \psi, \quad  \psi_t = B\psi,
\end{equation}
where $B=B(u,u_x, u_{xx},\ldots)$ is another linear operator with coefficients depending on $u$ and its derivatives.  Isospectrality implies a compatibility condition between $L$ and $B$, namely
\begin{equation}\label{eq:classical-compatibility}
L_t - [B, L] = 0.
\end{equation}
Under the appropriate choices\footnote{It is a long-standing open problem to characterize the PDEs which admit a Lax formulation.}  
 of $L, B$, this equation agrees with the nonlinear PDE of interest, and the operators $L$ and $B$ are called a Lax Pair for the PDE \cite{Lax68}.
\begin{eg}
One of the first equations to be successfully studied in this setting is the Korteweg-de Vries or KdV equation
\begin{equation}\label{eq:KdV}
u_t-6uu_x+u_{xxx} = 0,
\end{equation}
first used in modeling shallow water wave dynamics.  Defining the operators $L=-\frac{d^2}{dx^2} + u$ and $B=-4\frac{d^3}{dx^3} + 6u\frac{d}{dx} +3 u_x$, it is easy to verify that the compatibility condition \eqref{eq:classical-compatibility} is satisfied if and only if $u$ satisfies \eqref{eq:KdV}.  Notice that in this case the first equation in \refeq{eq:LB} is simply a classical eigenvalue problem for a linear Schr\"odinger operator having $u$ as potential.

\end{eg}
As the above example demonstrates, the operator $B$ depends on the unknown $u$, and may indeed be considerably more complicated than $L$, so solving the Lax system \refeq{eq:LB} may not be a straightforward task, except possibly at $t=0$.   However, the appearance in \refeq{eq:LB} of time-evolution for the eigenfunction $\psi$, provides a starting point for the Inverse Scattering Method (ISM), a nonlinear analogue of the Fourier Transform, to be applicable.  The three steps of the ISM are depicted in Figure \ref{fig:ISM} (excerpted from \cite{DrazinJohnson}), in which evolution of the scattering data determined by the eigenvalue problem is used to recover a potential $u(t,x)$ from the Cauchy data $u(0,x)$ of the nonlinear PDE of interest.
\begin{figure}[h]
\[
\xymatrixcolsep{5pc}
\xymatrix{
u(0,x)	\ar@{-->}[d] \ar[r]^{Dir\,\,Scattering}_{L \psi = \lambda \psi}	& S(0,\lambda) \ar[d]^{\psi_t = B \psi} \\ 
u(t,x)								& S(t,\lambda) \ar[l]^{Inv\,\,Scattering}_{G-L-M}	
}
\]
\caption{Classical ISM}\label{fig:ISM}
\end{figure}

In the first step, direct scattering refers to finding a matrix $S(0,\la)$ relating asymptotic eigenstates at $x=\pm \infty$ by way of ${\displaystyle \lim_{x \rightarrow \infty}\psi(0,x) = S(0 ,\lambda) \lim_{x \rightarrow -\infty}\psi(0,x)}$. Note that $S$ is a matrix since the asymptotic eigenspaces are multidimensional. This step requires identifying a Lax pair\footnote{It also hinges on space being non-compact and one dimensional.}.  For integrable equations, isospectrality of the operator $L$ implies that the time evolution of $S$ is linear; consequently, in the second step, the scattering matrix $S(t,\lambda)$ can be found (rightmost arrow).  Upon integration of the Gelfand-Levitan-Marchenko equations associated with the system, the third step, inverse scattering, recovers a time-evolved solution to the original nonlinear PDE of interest.  Standard techniques are described in \cite{GGKM, AblowitzClarkson, AKNS-1} and references therein.

For what follows, we will be careful to distinguish between integrability of a PDE (or system of PDEs) and existence of a solution-generating mechanism for that PDE, noting that many of the examples of this discussion will possess both features.

\subsection{Classical integrability, extended: the sine-Gordon equation}

Expanding the scope of classical integrability to address second order evolution equations involves going from scalar  linear operators to matrix operators, sometimes at the expense of introducing powers of $\la$, the spectral parameter\footnote{Although one is no longer considering an eigenvalue problem in the classical sense, $\la$ is often still called a spectral parameter.}.  In this case, matrix equations comparable to \eqref{eq:LB} are given by
\begin{equation}\label{eq:UV-system}
\partial_x \psi = U \psi, \quad \partial_t \psi = V \psi,
\end{equation}
where $\psi(t,x,\la):\RR^2 \times \CC \rightarrow \CC^{2\times2}$ and $U,V$ are $2\times 2$ matrix functions depending on $u,\la$.  The compatibility or zero-curvature condition corresponding to \eqref{eq:classical-compatibility} is  
\begin{equation}\label{eq:matrix-compatibility}
U_t-V_x+[U,V]=0.
\end{equation}
We take a moment to justify use of the phrase ``zero-curvature" by describing the geometry afforded by \eqref{eq:UV-system} and \eqref{eq:matrix-compatibility}.  Observe that \eqref{eq:UV-system} can be used to define a connection on a two-dimensional vector bundle over $\RR^2$ by setting
\begin{equation}\label{eq:nabla}
\nabla_{\partial_x}\psi = (\partial_x - U) \psi , \quad \nabla_{\partial_t}\psi = (\partial_t - V)\psi,
\end{equation}
and extending linearly to $\nabla_X$, for a vector field $X$.  In this manner, we may view the two equations in \eqref{eq:nabla} as parallel transport of $\psi$ in the $x$ and $t$ directions, respectively.  It is then straightforward to calculate the curvature of this connection for the basis vector fields $X=\partial_x ,Y= \partial_t$ to be
\begin{equation}
\nabla_X\nabla_Y \psi - \nabla_Y\nabla_X \psi - \nabla_{[X,Y]}\psi = ( U_t -V_x + [U,V] ) \psi.
\end{equation}
Thus, the connection has zero curvature if parallel transport of $\psi$ along any path connecting two points agrees, i.e.,  precisely if compatibility condition \eqref{eq:matrix-compatibility} is satisfied.
\begin{eg}\label{eg:sG} 
Matrix formulation of the ISM is used to address the integrability of the {\em sine-Gordon equation} (in null coordinates)
\begin{equation}\label{eq:sine-Gordon}
u_{\zeta\eta} = \sin u,
\end{equation}
first used in the study of mechanical solitons and differential geometry of constant curvature surfaces\footnote{Using the transformation $x=\zeta +\eta, t=\zeta-\eta$, one may easily recover the second recognizable form of the sine-Gordon equation, $u_{xx}-u_{tt} = \sin u$.}.  Defining $U,V$ as
\[
U = -i \lambda \begin{pmatrix} 1 & 0 \\ 0 & -1 \end{pmatrix} + \frac{1}{2} \begin{pmatrix} 0 & u_\zeta \\ u_\zeta & 0 \end{pmatrix} \quad\mbox{and}\quad V = \frac{i}{4\lambda} \begin{pmatrix} \cos u & \sin u \\  \sin u & -\cos u \end{pmatrix},
\]
the Lax system analogous to \refeq{eq:UV-system} is given by $\partial_\zeta \psi = U \psi$, $\partial_\eta \psi = V \psi$ and the zero curvature condition $U_\eta-V_\zeta+[U,V]=0$ ensures that $u(\zeta,\eta)$ evolves according to  \eqref{eq:sine-Gordon}.  A comparable initial value problem to the one depicted in Figure \ref{fig:ISM} is explicitly solved for the sine-Gordon equation by the ISM in 
 \cite{AKNS-3}; full details for the inverse scattering procedure may be found in \cite{ZS-I, ZS-II}.  It is worth noting that various Lax formulations are possible, by choosing different $U,V$ in \eqref{eq:UV-system} (see, for instance,  \cite{SG}).
\end{eg}

\subsection{From sine-Gordon to Chiral Fields}

Progression from matrix Lax pairs for a single nonlinear PDE to matrix Lax pairs for nonlinear {\em systems} of PDEs occurs naturally when studying simple geometric field theories, as the next example demonstrates.  What is surprising is that from the appropriate perspective, this matrix operator modification applies equally well to equations of gravitation and electrodynamics.
\begin{eg}\label{eg:cF}
Let $U$ be an open subset of $\RR^2$ with coordinates $(\zeta, \eta)$ and suppose $g: U \rightarrow G$ is a smooth function of $(\zeta, \eta)$ with values in a semisimple Lie group $G \subseteq GL_n(\mathbb{C})$.  Denote $(g^{-1})_\eta$ by $g^{-1}_\eta$ and consider the following action
\begin{equation}\label{eq:chiral-action}
S[g] = \int_U    \frac{1}{2} \mbox{Tr }\left( g_\zeta g^{-1}_\eta \right)  \, d\zeta d\eta.  
\end{equation}

The Euler-Lagrange equations of this action are
known as equations of the \emph{principal chiral field} on the group $G$--a free field in two-dimensional spacetime with values in $G$ \cite{ZM, ZS-I, Uhl89}. 
The word {\em chiral} here refers to the fact that this Lagrangian is invariant separately under the left, and the right, action of the group $G$ on itself.  With a non-commutative group, this action can thus be used to model physical phenomena that appear to break the chiral symmetry, such as the interaction of mesons in the chiral limit \cite{Gur60}.  In that case $G = SU(N)$ where $N$ is the number of quark flavors.

By introducing the compact notation $A=- g_\zeta g^{-1}$, $B=g_\eta g^{-1}$, and observing that $g^{-1}_\eta = - g^{-1}g_\eta g^{-1}$, the Lagrangian density in \eqref{eq:chiral-action} becomes $\frac{1}{2} \mbox{Tr} AB$.  Consequently, the field equations $(g_\zeta g^{-1})_\eta + (g_\eta g^{-1})_\zeta = 0$ are re-expressed as
\begin{eqnarray}\label{eq:Chiral-AB-System}
A_\eta - B_\zeta &=& 0 \\
A_\eta + B_\zeta +  [A,B] &=& 0, \nonumber
\end{eqnarray}
where the second equation is the compatibility condition arising from the definitions of $A$ and $B$.

To solve these equations for $g$ by means of inverse scattering, we cast \eqref{eq:Chiral-AB-System} as the compatibility condition of a linear system of matrix equations of the form
\begin{eqnarray}\label{eq:Chiral-Psi}
\partial_\zeta \psi = U \psi, \quad \partial_\eta \psi = V \psi,
\end{eqnarray}
for $\psi$ a vector function of $\zeta , \eta$ and $U,V$ $n\times n$ matrix functions of $\zeta, \eta$ and a complex parameter $\lambda \in \mathbb{C}$.  Equality of mixed partials in $\psi$ imposes the compatibility condition just as in \eqref{eq:matrix-compatibility}, replacing $x$ and $t$ by $\zeta$ and $\eta$, respectively, yielding
\begin{equation}\label{eq:UV-Compatibility}
U_\eta - V_\zeta + [U,V] =0.
\end{equation}

There are several quantities to determine in \refeq{eq:Chiral-Psi}, with varying degrees of freedom.  The matrices $U,V$ are to be determined in terms of $\zeta, \eta$ and $\lambda$, bearing in mind the compatibility conditions of both $U,V$ and $A,B$; furthermore, given a solutions $\psi$ of the linear system, the desired matrix $g$ should be recoverable by fixing a value of the parameter $\lambda$, under the definitions of $A$ and $B$.  Zakharov and Shabat carried out a \emph{vesture method} (later called the \emph{dressing technique}) to solve this overdetermined system of conditions by assuming $U,V$ have simple poles in $\lambda$:
\begin{equation}\label{eq:UV-Assumption}
U = U_0 + \sum_{n=1}^{N_1} \frac{U_n}{\lambda - \lambda_n}, \qquad \qquad
V = V_0 + \sum_{n=1}^{N_2} \frac{V_n}{\lambda - \mu_n}. 
\end{equation}
Substituting these expressions into zero-curvature condition \eqref{eq:UV-Compatibility} and clearing denominators, one obtains a polynomial in $\la$  of degree $N_1+N_2$, which splits into $N_1+N_2+1$ equations corresponding to coefficients of the polynomial \cite{ZS-I, ZS-II}.  

In the special case $N_1=N_2=1$, where $U,V$ each have one simple pole the compatibility condition \eqref{eq:UV-Compatibility} reduces to a degree 2 polynomial in $\lambda$ from which three equations are obtained.  For instance, assuming $\lambda_1=- \mu_1=1$ and under appropriate assumptions on $U_j, V_j$, \eqref{eq:Chiral-Psi} can be reduced to a Lax formulation of sine-Gordon equivalent to the one given in Example \ref{eg:sG}.  A complete analysis appears in  \cite{SG, FaddeevTakhtajan}.  Note that just as in the previous examples, \eqref{eq:Chiral-Psi} naturally defines the covariant differentiation operations given by $\partial_\zeta - \frac{A}{1-\la}$ and $\partial_\eta-\frac{B}{1+\la}$.
\end{eg}
\begin{rem}
Under the coordinate transformation $\zeta = \frac{1}{2}(t+x), \eta = \frac{1}{2}(t-x)$, the Lagrangian density of the action \eqref{eq:chiral-action} differs by a null Lagrangian from $\frac{1}{2} \mbox{Tr} \, (\tilde{B}^2 -  \tilde{A}^2)$, where $\tilde{A}= g_tg^{-1}$ and $\tilde{B}=g_x g^{-1}$.  Noting that $\tilde{A},\tilde{B}$  must lie in the Lie algebra $\mathfrak{g}$ of the Lie group $G$, the transformed action may be viewed as an inner product of the matrix-valued one-form $W=\tilde{A}dt + \tilde{B}dx$ with itself, relative to the Minkowski metric on $\RR^2$.
As we shall see, this one form can be interpreted as the pull-back of the Maurer-Cartan form $w=-dg g^{-1}$ under the mapping $g(x,t)$.  
We shall also see below that the principal chiral field is simply a harmonic map into a Lie group.  These two facts will combine in a particularly elegant manner where the integrability of such maps is studied.
\end{rem}



\section{Gravitational Equations}

Motivated by the last example, we turn our attention to the Einstein vacuum equations with vanishing cosmological constant
\beq\label{eq:EVE}
\mathbf{R}_{\mu\nu} =0, \qquad \mu, \nu = 0, \ldots 3.
\eeq
This quasilinear system is satisfied by the metric tensor of a four-dimensional Lorentian manifold $(\cM, \bg)$.  Here, $\mathbf{R}_{\mu\nu}$ denotes the Ricci curvature tensor of $\bg$.  Complexity of these field equations is reduced considerably using symmetry reductions, corresponding to existence of Killing fields for the metric $\bg$.  In this context, existence of timelike Killing fields correspond to  stationary metrics, and existence of spacelike Killing fields correspond to spherically symmetric or axisymmetric metrics (depending on whether the action generated is an $SO(3)$ or $SO(2)$ rotation, respectively).  Under such symmetries, the simplest nontrivial asymptotically flat solutions\footnote{Asymptotically flat solutions represent vacuum outside an isolated body.} are Schwarzschild spacetimes, which are static and spherically symmetric, and the Kerr spacetimes, which are stationary and axisymmetric  \cite{Schw, Kerr63}.

We shall describe two approaches for the analysis of a two-Killing field reduction of \eqref{eq:EVE}, namely the Zakharov-Belinski and the Ernst formulations, and establish a concrete connection between them by way of harmonic maps.

\subsection{Integrable formulation of the Einstein vacuum equations}
First, we briefly outline the techniques in \cite{ZB-I, ZB-II} used to establish integrability of the stationary, axisymmetric Einstein vacuum equations.  Assuming the existence of two commuting Killing fields, the spacetime metric $\bg$ can  be expressed in block-diagonal form
\beq
\bg_{\mu\nu}dx^{\mu}dx^{\nu} = f(\rho,z)(d\rho^2 \pm dz^2)+\tilde{\bg}_{ab}(y,z)dx^adx^b \quad a,b=1,2,
\eeq
with $x^0=\rho$ and $x^3=z$ \cite{Papapetrou}.  Here the $+$ sign is to be used when both Killing fields are spacelike (and hence $\tilde{\bg}$ is Lorentzian), while the $-$ sign is for the case of one timelike  and one spacelike Killing field (in which case the quotient metric $\tilde{\bg}$ is Riemannian).  In the former case, let $\zeta = \rho +z, \eta = \rho -z$ be null coordinates on the quotient manifold, and in the latter, set $\zeta = \rho+iz, \eta = \bar{\zeta}$.  The equations for $\tilde{\bg}$ are given by
\begin{eqnarray}\label{eq:EVE-AB-system}
A_\eta - B_\zeta &=& 0 \\
A_\eta + B_\zeta + \frac{1}{\alpha} [A,B] - \frac{\alpha_\eta}{\alpha}A + \frac{\alpha_\zeta}{\alpha}B &=& 0, \nonumber
\end{eqnarray}
for $\det \tilde{\bg} = \alpha ^2$,  $A= -\alpha \tilde{\bg}_\zeta \tilde{\bg}^{-1}$, $B = \alpha \tilde{\bg}_\eta \tilde{\bg}^{-1}$.  Compare with the chiral field model \eqref{eq:Chiral-AB-System}.  The task is to find a linear system for $U,V$ analogous to \eqref{eq:Chiral-Psi} for which the above PDE system appears as a compatibility condition; note that once $\tilde{\bg}$ is determined, $f$ can be found by quadrature (see, e.g. \cite{Wei94}).

Just as in \eqref{eq:UV-system} for the sine-Gordon equation and  \eqref{eq:Chiral-Psi} for the chiral field, assume $U$ and $V$ have simple poles in $\la$.  The key idea used in establishing integrability in this case is a generalization of  $\partial_j$ to operators $D_j$ which include differentiation in the \emph{spectral parameter} $\la$, namely
\begin{equation}\label{eq:ZB-Psi}
D_1\psi  = U\psi := \frac{A}{\lambda - \alpha} \psi,  \quad D_2\psi = V\psi := \frac{B}{\lambda + \alpha} \psi,
\end{equation}
where $D_j = \partial_j - p_j\partial_\lambda$, and $p_j= p_j(\zeta,\eta, \lambda)$ is a rational function of $\la$.  We shall interpret these generalized derivatives geometrically in Section 4. 
Note that here $\al = \al(\rho,z)$, thus in contrast to the chiral field model, the poles in \refeq{eq:ZB-Psi} are not fixed, but {\em moving}.

With integrability established, the authors of \cite{ZB-I, ZB-II} implement a vesture or dressing method to generate new solutions to the Einstein vacuum equations.   Formally, given a solution $\psi_0$ of the Lax system having initial data $\bg_0$, dressing refers to the procedure of finding a matrix $\chi$ for which $\psi = \chi \psi_0$ also solves the system.  It is from this ``dressed" generating matrix $\psi$ that new solutions $\tilde{\bg}$ (and hence $\bg$) are constructed.  Determining the dressing matrix $\chi$ reduces to solving a finite system of algebraic equations.  In the next section we shall describe a generalization of this technique in the context of dressing harmonic maps.

\subsection{Ernst Formulation of the vacuum equations}
Prior to the works of \cite{ZB-I, ZB-II}, another approach to the two-Killing field reduction of EVE had been investigated in \cite{Ern68a, Ern68b}, giving what is now known as the Ernst Equation for a complex potential function $\varepsilon=\varepsilon(\rho,z)$.  We briefly review this now.

Let the two commuting Killing fields for the metric $\bg$ be denoted by $\bK = \frac{\p}{\p t}$ and $\tilde{\bK} = \frac{\p}{\p \varphi}$.  Let us assume that $\bK$ is timelike and $\tilde{\bK}$ spacelike.  Set 
\beq
X := - \bg(\bK,\bK)>0, \ \tilde{X} = \bg(\tilde{\bK},\tilde{\bK}),\ W := \bg(\bK,\tilde{\bK}),\  \Om := \frac{W}{X},
\eeq
and define
\beq
 \rho := \sqrt{W^2 + X\tilde{X}},\qquad  \bb := i_\bK *d\bK.
\eeq
Thus, $\rho$ is the element of area of the cylindrical group orbits and $\bb$ is the {\em twist form} of $\bK$.  The line element of $\bg$ is then
\begin{eqnarray}
ds_\bg^2 &=& -X dt^2 + 2W dt d\varphi + \tilde{X} d\varphi^2 + \tilde{\bg}_{ab} dy^a dy^b \\
&=& -X(dt+ \Om d\varphi)^2 + \frac{1}{X} (\rho^2d\varphi^2 + ds_\bm^2), \nonumber
\end{eqnarray}
where $\bm$ is a metric on the two-dimensional Riemannian quotient manifold conformal to $\tilde{\bg}$.  

Suppose now that the metric $\bg$ is a solution of Einstein vacuum equations. It then follows that (i) the function $\rho$ is harmonic $\Delta_\bm \rho = 0$, and  (ii) $\bb$ is closed, i.e. $d\bb = 0$.  Thus (assuming the domain is simply connected) $\bb = dY$. The function $Y$ is called the {\em twist potential} for the Killing field $\bK$ (see e.g. \cite{Wei94}).  Choosing a conjugate harmonic function $z$ for $\rho$, and using $(\rho,z)$ as {\em isothermal} coordinates on the quotient brings the line element to the form
\beq\label{eq:LewPap} ds_\bg^2 = -X(dt+\Om d\varphi)^2 + \frac{1}{X} (\rho^2 d\varphi^2 + e^{2u}(d\rho^2 + dz^2)).
\eeq
This is known as the Lewis-Papapetrou form of the metric.  There are now only three unknown metric coefficients left: $X$, $\Om$ and $u$. Next, by virtue of $\bK$ and $\tilde{\bK}$ being Killing fields (and the Einstein vacuum equations being satisfied), it is possible to show that the quantities $X$ and $Y$ satisfy a coupled system of elliptic PDEs, which can be combined into a single equation by introducing a complex-valued potential $\varepsilon = X+iY$, called the {\em Ernst potential}:
\beq\label{eq:Ernst}
(\varepsilon + \bar{\varepsilon})\triangle \varepsilon + 2 \nabla \varepsilon \cdot \nabla \varepsilon = 0.
\eeq
Once the potential $\varepsilon$ is found, the remaining metric coefficients $\Om$ and $u$ can be found using quadratures \cite{Wei94}.

Equation \refeq{eq:Ernst} is derivable from a variational principle, namely, it is the Euler-Lagrange equation of the following action
\beq\label{eq:Ernst-Action}
\int_{\RR^2} \frac{1}{(\Re \varepsilon)^2} |\nabla \varepsilon|^2 d\bx =  \int\int \frac{1}{4X^2}\left( |\nabla X|^2 + |\nabla Y|^2\right) \rho \, d\rho dz
\eeq
We will see below that this is also the action for a harmonic map into the hyperbolic plane. 

\subsection{Harmonic Maps and Gravitation}
Recall that a harmonic map $f:(\cM, \bg) \to (\cN, \bh)$ is a critical point of the Dirichlet Energy
\beq\label{eq:Harmonic-Action}
\cE[f,\cD] = \int_\cD \half \tr_\bg f^*\bh.
\eeq
This definition generalizes the notion of a geodesic to higher-dimensional domains: letting $(\cM, \bg) = (\RR^1, id)$, the Euler-Lagrange equations for the harmonic map action reduce precisely to the equations of an energy-minimizing geodesic on $\cN$.  Harmonic maps also generalize harmonic functions to nonlinear targets: letting $(\cN,\bh)=(\RR^1, id)$, the Euler-Lagrange equation for \refeq{eq:Harmonic-Action} corresponds to that of a harmonic function on the base manifold $\cM$.  With this notation in place, we can cast some of our prior examples from the previous sections in the context of harmonic maps\footnote{The sine-Gordon equation can also be put into this framework, see \cite{ShatStrauss96}.}.

{\em Example \ref{eg:cF}, revisited.}
To express the chiral field appearing in Example \ref{eg:cF} as a {\em wave map}, the hyperbolic analog of a harmonic map, let $g = f(\eta,\zeta)$ with $f :\RR^{1,1} \rightarrow G$ a mapping from the $1+1$-dimensional Minkowski space into the group $G$,  and $(\eta,\zeta)$ null coordinates on the domain. The Maurer-Cartan form of $G$ is $w = -dg g^{-1}$, a Lie-algebra-valued 1-form.  Let $W$ denote its pull-back under $f$.  Thus $W = \theta^aX_a$ where $\{X_a\}$ is a basis for the Lie algebra $\fg$ of the group $G$, and  $\theta^a$ are 1-forms on the domain: $\theta^a = \theta^a_\mu dx^\mu$.  On the other hand, if $G$ is semi-simple, the Killing-Cartan quadratic form $\kappa_{ab} := \half \tr (X_a X_b)$ is non-degenerate, and it thus endows $G$ with a pseudo-Riemannian\footnote{The metric is Riemannian if $G$ is a compact group.} metric $\bh = \half \tr (w^2)$.  With $\bg$ denoting the Minkowski metric, we thus have
\beq
\half \tr_\bg f^*\bh  = \half \bg^{\mu\nu} \kappa_{ab} \theta^a_\mu \theta^b_\nu = \half \bg^{\mu\nu}\tr (X_aX_b)\theta^a_\mu \theta^b_\nu  = \half \tr (W^2),
\eeq
which agrees with \eqref{eq:chiral-action}.  More precisely, $W = (-g_\zeta g^{-1}, - g_\eta g^{-1})$, so that $\tr(W^2) = \tr (g_\zeta g^{-1}_\eta)$, the chiral field Lagrangian, as before. 

We now move on to the reduction of vacuum Einstein equations under the assumption of existence of two commuting Killing fields.
\begin{eg}\label{eg:EVE}
Let $(\cM,\bg)$ denote a solution of Einstein's vacuum equations possessing two commuting Killing fields.  We may then express the line element of $\bg$ in the Lewis-Papapetrou form \refeq{eq:LewPap}.  Let $(X,Y) \in \RR^+\times \RR$ be coordinates on the real hyperbolic plane $\cN = \Hset_\RR$ with its standard metric $\bh$,
\beq\label{eq:hyperbolic-metric}
ds_\bh^2 = \frac{dX^2+dY^2}{X^2}.
\eeq
Suppose  $f: (\cM,\bg) \to (\cN,\bh)$ is a harmonic map that is invariant under the action of the two given Killing fields, so that $f$ is well-defined on the quotient manifold, $f = (X(\rho,z),Y(\rho,z))$.  The harmonic map Lagrangian density is therefore
\begin{eqnarray}
 \tr_\bg f^*\bh \sqrt{-\det \bg}  &=&  \frac{1}{4X^2}\left\{g^{\rho\rho}[(\p_\rho X)^2 + (\p_\rho Y)^2] + g^{zz}[(\p_z X)^2 + (\p_z Y)^2]\right\} \frac{\rho e^{2u}}{X} \nonumber\\
& = &  \frac{1}{4X^2}\left( |\nabla X|^2 + |\nabla Y|^2\right) \rho
\end{eqnarray}
which is the same as \eqref{eq:Ernst-Action}, the Lagrangian obtained for the Ernst equation.  
Indeed, since the action is independent of all other features of the spacetime metric, we can pretend that the domain is simply $\RR^3$ in cylindrical coordinates $(\rho,z,\varphi)$, and that we have an axisymmetric (i.e. $\varphi$-independent) harmonic map from $\mathbb{R}^3$ into the hyperbolic plane.

Finally, observing that the real hyperbolic plane may be realized as a symmetric space $\mathbb{H}_{\mathbb{R}} \cong SL(2, \mathbb{R})/SO(2)$,
we may use the Cartan embedding of a symmetric space $G/K$ into its mother Lie group $G$, to represent $(X,Y)$ as a mapping into $SL(2,\RR)$, namely
\beq\label{eq:Xdef}
(X,Y) \to \frac{1}{X} \left(\begin{array}{cc} X^2+Y^2 & Y \\ Y & 1\end{array}\right).
\eeq
The fact that the target of the map is now embedded in a linear space\footnote{Note that by virtue of the Cartan embedding, a symmetric space can be embedded into a Lie group, for which one may choose a matrix representation, thus making addition and scalar multiplication of elements possible.} is the key to the success of the vesture method, described in the next section.
\end{eg}
\begin{eg}\label{eg:EMM}
The Einstein-Maxwell system, equations governing the interaction of the spacetime metric $\bg$ with an electromagnetic field $\bF$, given by
 \begin{equation}\label{eq:Einstein-Maxwell}
 \mathbf{R}_{\mu\nu} - \half \mathbf{g}_{\mu\nu}R = \kappa \mathbf{T}_{\mu\nu}, \quad \mathbf{T}_{\mu\nu} := \bF_\mu^\la *\bF_{\nu\la} - \bg_{\mu\nu} \bF_{\alpha\beta}\bF^{\alpha\beta}, \quad d\bF = 0, \quad d*\bF = 0,
 \end{equation}
also represent the equations of an axially symmetric harmonic map, assuming existence of two commuting Killing fields which also leave the field $\bF$ invariant.  In this case, the mapping is given by $f:\cM \to \cN$, where $\cM$ is $\RR_+^2 \times \mathbb{S}^1$ as in the previous example, and $\cN$ is the {\em complex} hyperbolic plane realized as $\Hset_\Cset = G/K$  with $G=SU(2,1)$, and $K=S(U(2)\times U(1))$ \cite{Ern68b, Car73, Maz84}, and the corresponding Ernst formulation now involves a {\em pair} of complex potentials $(\varepsilon, \Phi)$ \cite{Ern68a, Ern68b, Chandrasekhar-1}.
\end{eg}

Appearances of harmonic maps in physics have been extensively documented, as has the integrability of harmonic maps into particular targets; see \cite{Misner, NSanch82, Woo94}, for instance.  The primary observation we now make is the following:  the chiral field, Einstein vacuum and Einstein-Maxwell equations (under a two-Killing-field reduction) are integrable and admit a dressing mechanism simply because \emph{harmonic maps into symmetric spaces are integrable and admit a generalized dressing mechanism}.

\section{Integrability and vesture of axially symmetric harmonic maps}

Motivated by previous work, we prove a pair of theorems in \cite{SB-STZ} which unify the integrability and solution-generating mechanism (vesture or dressing technique) of special cases considered in the literature for the Einstein vacuum, Einstein-Maxwell and Chiral Field models; the results further establish integrability and vesture for a broader class of geometric field theories.  We shall outline the main points surrounding their proof, after which the second half of this communication will focus on control of the solution-generating mechanism.

\subsection{Integrability of axially symmetric harmonic maps}

\begin{thm}\label{thm:integrability} (Integrability)
Let $G$ be a real semisimple Lie group and $K$ a maximal compact subgroup.  Then any axially symmetric harmonic map from $\mathbb{R}^3$ into the Riemannian symmetric space $G/K$ satisfies an integrable system of equations.
\end{thm}
The proof of this statement involves two main steps.  Let $(\cM, \bg) =( \RR^3, \bg)$ where the metric is given in cylindrical coordinates $(\rho,z,\varphi)\in \RR^{2}_+ \times \Sset^1$.  Suppose $f:\cM \to G/K$ is an axially symmetric harmonic map and let $q$ be a parametrization of the symmetric space via the Cartan embedding $G/K \to G$, so that $q = f(\rho,z) \in G$.

\emph{Step $a.$ Rewrite the equations of a harmonic map as a Hodge System.}  Using the pull-back of the Maurer-Cartan form $W = dq q^{-1} \in \bigwedge^1(\cM)$, the harmonic map equations are
\beq\label{eq:W}
dW + W\wedge W = 0,\qquad d(\rho *W) = 0,
\eeq
where the domain is the right-half plane in $\RR^2$ with coordinates $\bx =(\rho,z)$. 

\emph{Step $b.$ Exhibit a Lax Pair whose compatibility condition is the Hodge System.}  In our case, we generalize the linear operators $D_j$ appearing in \eqref{eq:ZB-Psi} by defining 
\begin{equation} \label{eq:Psisys}
D\Psi  =  - \Omega \Psi, \qquad \Psi |_{\la = 0} =  q,
\end{equation}
\begin{equation*}
D_\mu := \p_\mu - \om_\mu\frac{\p}{\p \la}, \quad  \om_\mu := \frac{1}{\p\varpi/\p\la }\p_\mu\varpi , \quad  \Omega_\mu := a W_\mu + b \rho (*W)_\mu,\qquad \mu = 1,2, \nonumber
\end{equation*}
where $\la \in \bar{\Cset}$, $\bx =(\rho,z)$, and $\varpi(\la,\bx)$, $a(\la,\bx)$, $b(\la,\bx)$ are three specific $\bar{\Cset}$-valued functions on $\bar{\Cset} \times \RR^2_+$ defined to be
\beq
 \varpi(\la,\bx) = \frac{\rho^2}{2\la} + z - \frac{\la}{2}, \quad  a(\la,\bx) = \frac{\rho^2}{\la^2 + \rho^2},\quad b(\la,\bx) = \frac{\la}{\la^2+\rho^2}.
\eeq

The unknown is $\Psi: \Cset\times \RR^2_+ \to \Cset^{n\times n}$, solving the Lax System \eqref{eq:Psisys}. Here $n$ is the dimension of a regular representation of the Lie group $G \subset GL(n,\RR)$.   Note that this treatment subsumes approaches appearing in \cite{ZB-I, ZB-II, Ale81, EGK84}.
\begin{rem}
We remark that just as in Examples \ref{eg:sG}, \ref{eg:cF}, the operator $D$ defines a connection on the domain of $\Psi$, a Riemann surface bundle $\displaystyle{ \cB=\!\!\bigcup_{\bx \in\RR^2_+}\!\!\cR_\bx}$, where $\cR_\bx$ is the two-sheeted Riemann surface defined by 
\begin{equation}\label{eq:RiemannSurface}
\cR_\bx = \{(\la,\varpi)\in \Cset^2\ |\  \la^2 -2\la (z-\varpi) - \rho^2 = 0\}.
\end{equation}
The mapping $T_\bx:\bar{\Cset} \to \bar{\Cset}$, $T_\bx(\la) = -\rho^2/\la$ is a deck transformation on the universal cover of $\cR_\bx$ and $T_\bx^2=id$ for all $\bx \in \RR^2_+$.  We prove in \cite{SB-STZ} that the compatibility condition of the $D_\mu$ is equivalent to a zero curvature condition $d\Om + \Om \wedge \Om=0$, for the 1-form $\Omega$ on the domain (cf.  \cite{Guest08}, p. 54).
\end{rem}

\subsection{Vesture of axially symmetric harmonic maps}

\begin{thm}\label{thm:vesture} (Vesture)
If the Lie group $G$ is such that the two involutions $\tau$ and $\sigma$ defining the symmetric space $G/K$ can be given by conjugation with the same element, then the vesture method can be used to construct new harmonic maps starting from any given harmonic map.
\end{thm}
As stated earlier, vesture refers to the procedure of finding a dressing matrix $\chi(\bx , \la)$ such that if $\Psi_0$ is a solution of \refeq{eq:Psisys} and
\begin{equation}\label{eq:chidressing}
\Psi=\chi\Psi_0,
\end{equation}
then $\Psi$ also solves the Lax system \eqref{eq:Psisys}.  Using the notation set up in Theorem \ref{thm:integrability}, we prove this result in two main steps as well.

\emph{Step $c.$ Construct a dressing matrix $\chi$ possessing the appropriate symmetries.}  An invertible mapping $\chi:\overline{\Cset}\times \RR^2_+ \to H$ is a {\em dressing matrix} for $q_0$, an axially symmetric harmonic map of $\RR^3$ into $G/K$, if it satisfies
\begin{eqnarray}\label{eq:chi-symmetries}
\chi(\infty) &=& I \nonumber \\ 
\tau(\chi(\overline{\la})) &=& \chi(\la) \\
\chi(\la) &=& \chi(0)q_0 \sigma (\chi(T_\bx(\la)))\sigma( q_0). \nonumber
\end{eqnarray}
Here, $\tau, \sigma$ are commuting involutions characterizing the symmetric space $G/K$ and $H$ is a complex group containing $G$ \cite{Helgason-1}; the second and third equations in \eqref{eq:chi-symmetries} are referred to as the G-reality and involutive symmetry conditions, movitated by the study of wave maps into compact symmetric spaces in \cite{TerUhl04, Terng10}.  The symmetry conditions in \eqref{eq:chi-symmetries} ensure that the resulting axially symmetric harmonic map $q(\bx) = \chi(0,\bx) q_0(\bx)$, found by setting $\la=0$ in the generating matrix $\Psi$, takes its values in the appropriate target space, $G/K$.  We show that there always exists an element of the equivalence class of $\Psi$ which satisfies such imposed symmetries \cite{SB-STZ}.

\emph{Step $d.$ Reduce the problem of finding the dressing matrix to an algebraic problem.}  Motivated by simplifications appearing in \cite{ZB-I, ZB-II, EGK84}, we assume $\chi$ is a rational function of $\la$ and unknown matrices $R_k = R_k(\bx)$ of the form
\begin{equation}\label{eq:Chi-Ansatz}
\chi(\bx, \la) = I + \sum_{k=1}^{2N} \frac{R_k(\bx)}{\lambda - \la_k(\bx)}.
\end{equation}
The moving poles, $\la_k(\bx)$, are specified in the next section.  Under further assumptions on the involutions $\tau, \sigma$ we prove solvability of the linear system in $R_k$.

We summarize Theorems \ref{thm:integrability} and \ref{thm:vesture} in Figure \ref{fig:ismq}.  Steps $a.$ and $b.$ comprise the integrability component (top arrow)  and steps $c.$ and $d.$ comprise the vesture component (rightmost and bottom arrows) of the results.  Compare with Figure \ref{fig:ISM}.  In our case, the dressing procedure results in axisymmetric harmonic maps $q$ possessing ring singularities \cite{SB-STZ}.
\begin{figure}[h]
\[
\xymatrixcolsep{5pc}
\xymatrix{
q_0(\bx)	\ar@{-->}[d] \ar[r]^-{D\Psi=-\Omega \Psi}_-{\Psi |_{\lambda=0} \,= q_0}	&\Psi_0(q_0, \lambda) \ar[d]^{\Psi = \chi \Psi_0} \\
q(\bx)									& \Psi(q, \lambda) \ar[l]^-{\lambda = 0}_{}
}
\]
\caption{ISM for Harmonic Maps}\label{fig:ismq}
\end{figure}

The linear system resulting from this reduction unfortunately still contains too many unknown (and seemingly free) parameters.  Thus, the question posed at the beginning of this paper is restated as: Can we control the output of this mechanism to generate desired harmonic maps from a given initial seed map $q_0$, and if so, how?

\subsection{The Dressing Algorithm}
 
In order to control the number of parameters appearing in the dressing mechanism, we will need to outline the  solution-generating algorithm afforded by the proof of Theorem \ref{thm:vesture}.  We do so now.

Let $\{\varpi_k\}_{k=1}^N$ be $N$ distinct non-real complex numbers in the upper half plane.  For every $\bx = (\rho,z) \in \RR^2_+$, each $\varpi_k$ corresponds to a pair of points on the Riemann surface $\cR_x$ defined in \eqref{eq:RiemannSurface}.  Number these $2N$ prescribed  poles $\la_k (\bx)$ in such a way that they are related by the deck transformation on $\cR_\bx$:
\beq \label{lakom}
\la_{N+k} = T_\bx(\la_k)= \frac{-\rho^2}{\la_k},  \qquad 1\leq k \leq N.
\eeq
 Suppose that the involutions $\tau, \si$ characterizing the symmetric space target $G/K$ can be given by conjugation with respect to the same element  $\Ga$.  In conjunction with the G-reality and involutive symmetry conditions in \eqref{eq:chi-symmetries}, this assumption allows the algebraic reduction to be carried out efficiently\footnote{In particular the pseudo-unitary groups $SU(p,q)$ satisfy this assumption. Moreover, groups possessing Dynkin diagrams with no symmetries may also qualify.   Modification of this procedure to address other families of involutions will be taken up in a future work.}.

The overdetermined system in $R_k$ is further simplified using a rank-one assumption.  Define $M_k :=R_k(\bx)\Psi_{0k}(\bx)$, where $\Psi_{0k}(\bx) = \lim_{\la \rightarrow \la_k}\Psi_0(\la, \bx)$ and assume there exist non-zero vector functions $\bu_k,\bv_k$ for $k=1,\dots,2N$ such that
\beq\label{eq:rankone}
M_k = \bu_k \bv_k^*,
\eeq
and furthermore that the vectors $\bv_k$ satisfy 
\beq\label{eq:vk}
\bv_k = \Ga \bv_{N+k} \qquad  k = 1,\dots, N.
\eeq
This rank-one ansatz is consistent with the prior conditions imposed on the $R_k$ system by symmetry, and the relation on the vectors $\bv_k$ imposes a relation on the vectors $\bu_k$ which does not overdetermine the problem.

Although the $R_k$ system is equivalent to a nonlinear system for the $M_k$, it is shown in \cite{SB-STZ} that $\bv_k$ can be taken to be constant vectors, and that the rank one assumption reduces the $M_k$ system to a \emph{linear} system for the unknown vector functions $\bu_k$, given by
\begin{equation}\label{eq:aub}
\sum_j a_{kj} \bu^*_j = \bb_k^*,\qquad a_{kj} := \frac{1}{\la_k - \overline{\la}_j} \bv_k^*S_{kj}\bv_j,\qquad \bb_k^* := -\bv_k^* \Psi_{0k}^{-1} \Ga,
\end{equation}
where we have set 
\beq\label{def:Skj}
S_{kj} := \Psi_{0k}^{-1}\Ga (\Psi_{0j}^*)^{-1}=\Ga\Psi_{0k}^*\Ga \Psi_{0j} \Ga.
\eeq
Thus, a complete set of unknowns for the problem are the vectors $\{\bu_k\}_{k=1}^{2N}$, given an arbitrary set of constant vectors $\{\bv_k\}_{k=1}^N$ in $\CC^n$.  Equation \eqref{eq:aub} can be written more compactly in matrix form as  
\beq \label{linalg} 
AU^* = B^*.
\eeq
Here $U=U(\bx)$ is the $n\times 2N$ matrix whose columns are the vector functions $\bu_k$, $A = A(\bx) = (a_{kj})$ is a $2N\times 2N$ matrix function, and $B=B(\bx)$ is the $n\times 2N$ matrix whose columns are the vector functions $\bb_k$ defined above.

If the matrix $A$ can be shown to be invertible (at least in a neighborhood in $\RR^2_+$), then the above system has a unique solution $U^* = A^{-1}B^*$ in that neighborhood.  From there, one can then calculate the matrices $R_k$ and the dressing matrix $\chi$, and setting $\la =0$, the new solution $q$ is found to be
\beq\label{eq:solq}
q(\bx) = q_0(\bx) - \sum_{k=1}^{2N} \frac{1}{\la_k(\bx)} \bu_k(\bx) \bv_k^* \Psi_{0k}^{-1}(\bx) q_0(\bx).
\eeq

In general, $A$ is {\em not} invertible everywhere in the domain $\RR^2_+$, and indeed the zero set of $\det A$ has a geometric significance for the dressed harmonic map $q(\bx)$. 
 However, we can prove that under appropriate conditions on the arbitrary vectors $\bv_k$, the matrix $A(\bx)$ is, in general, invertible {\em for large }$|\bx|$ (i.e., in a neighborhood of infinity).  Once a $q(\bx)$ has been found, its domain of definition may be maximally extended by analytic continuation.
\begin{rem}\label{rem:scaleinvariance}
Observe that \eqref{eq:aub} and hence \eqref{eq:solq} satisfies scaling invariance:  if we rescale the constant vectors $\bv_k$ by (nonzero) complex constants $t_k \in \CC$, $k=1, \ldots N$, then the expressions for $a_{kj}$ also rescale by a factor of $t_k^*t_j$.  This suggests that by rescaling each $\bu_j$ by $\frac{1}{t_j^*}$, one preserves the equality in \eqref{eq:solq}.  Thus, we may view $\bu_j$ not as vectors in $\CC^{2N}$, rather as elements of $(\mathbb{PC})^{2N}$. This also means that the actual number of free real parameters in our dressing procedure is $nN$, and not $2nN$ as it initially appeared to be (from $N$ arbitrary constant vectors in $\CC^n$).  We shall invoke the scale invariance of the system explicitly when it is convenient to do so in our subsequent calculations.
\end{rem}

\section{$N$-solitonic solutions of Einstein's Equations}

The solution-generating algorithm outlined in the previous section is realized concretely for the case of the pseudo-unitary groups, $G=SU(p,q)$ in \cite{SB-STZ}.  Define involutions $\si, \tau$ on $H = SL(p+q, \Cset)$ by
\beq
\tau (g)= \Gamma g^{-*} \Gamma, \quad \sigma (g) = \Gamma g \Gamma, \quad 
\Gamma = \begin{bmatrix} I_{p\times p} & 0\\ 0 &-I_{q\times q} \end{bmatrix}.
\eeq
Then $G =\{ g \in SL(p+q,\Cset)  |  g^* \Gamma g = \Gamma \} $, $K =\{ g \in G\ | \ \sigma(g) = g\}$, and using the Cartan embedding, the symmetric space $G/K$ is the complex Grassmann manifold 
\beq
\cG_{p,q}:=G/K=SU(p,q)/S(U(p)\times U(q))  =\{ q \in G\ |\  q \Gamma q \Gamma = I \}.
\eeq
In the special case where $p= q=1$, one obtains  $G/K = \mathbb{H}_\mathbb{R}$, the target in the Ernst reduction of Einstein vacuum equations, appearing in Example \ref{eg:EVE}, and for the case $p=1, q=2$,  one obtains $G/K = \mathbb{H}_\mathbb{C}$, the target in the Ernst reduction of Einstein Maxwell equations, appearing in Example \ref{eg:EMM}.

\subsection{Kerr spacetimes from 1-soliton dressed Minkowski metrics}
We now specialize to the case $p=q=1$ to demonstrate the dressing algorithm explicitly, showing that dressing Minkowski spacetime with one soliton results in a family of (naked) Kerr spacetimes.
\begin{enumerate}
\item  (Choose initial seed) Recall from Example \ref{eg:EVE}, in the Ernst formulation of the vacuum Einstein equations, $\varepsilon  = X+ iY$, where as a harmonic map, $q = \frac{1}{X} \left( \begin{array}{cc} X^2+Y^2 & Y \\ Y & 1\end{array} \right)$.  Thus,  the Minkowski spacetime metric can be expressed as the harmonic map into $G/K$ given by
\beq
q_0(\bx)=I_{2\times 2}:\RR^2_+ \longrightarrow  SU(1,1)/S(U(1)\times U(1)),
\eeq
where $\bx=(\rho, z)$.  We solve the Lax system \eqref{eq:Psisys} by inspection and find $\Psi_0= I_{2\times 2}$.

\item  (Dress $q_0$) Fix $\varpi = i s$, $s>0$ and let $\la_1, \la_2$ be the two roots of
$p(\bx, \la):=\la^2 - 2(z-is)\la - \rho^2$ so that
\beq
 \chi(\bx, \la)  = I + \frac{R_1(\bx)}{\la-\overline{\la_1}}+ \frac{R_2(\bx)}{\la-\overline{\la_2}}.
\eeq

\item  (Linear Algebra) Set $\bv_1 = [\alpha \,\,\,\delta]^t$ for $\alpha, \delta \in \mathbb{C}$, so that by \eqref{eq:vk}, $\bv_2 = \Ga \bv_1 = [\alpha \,\,-\!\delta]^t$.  Rewrite $R_k$, as in \eqref{eq:aub}, in terms of $\alpha, \delta$ and vectors $\bu_k$:
\small{
\beq\label{eq:inversion}
\left[ \begin{array}{cc} \frac{1}{\la_1 - \overline{\la_1}}(|\alpha|^2-|\delta|^2) &  \frac{1}{\la_1 - \overline{\la_2}}(|\alpha|^2+|\delta|^2) \\ & \\ \frac{1}{\la_2 - \overline{\la_1}}(|\alpha|^2+|\delta|^2) &  \frac{1}{\la_2 - \overline{\la_2}}(|\alpha|^2-|\delta|^2) \end{array} \right] \left[ \begin{array}{c} \bu_1^*  \\ \\  \bu_2^*  \end{array} \right] 
= \left[ \begin{array}{cccc} -\overline{\alpha}  & \overline{\delta} \\ -\overline{\alpha}  & -\overline{\delta} \end{array} \right] 
=:  \left[ \begin{array}{c} -\bv_1^*\\-\bv_2^* \end{array} \right].
\eeq
}

Solving for $\bu_1, \bu_2$ above, the new harmonic map $q$ in \eqref{eq:solq}, expressed in terms of the complex constants $\alpha, \delta$ and $\la_j = \la_j(\bx)$, is given by
\beq\label{eq:1solitonsolq}
q(\bx) = \chi(\bx, 0)q_0(\bx)=I_{2\times2} - \frac{1}{\la_1}\bu_1(\bx)\bv_1^* - \frac{1}{\la_2}\bu_2(\bx)\bv_2^*.
\eeq

\item  (Change coordinates) We now change from Weyl coordinates $\bx=(\rho, z)$ to Boyer-Lindquist coordinates $(r, \theta)$, defined as
\beq
\rho = \sqrt{(r-m)^2+s^2} \sin \theta, \qquad z=(r-m)\cos \theta,
\eeq
for real parameters $m, s$. Then the harmonic map is re-expressed as
\small{
\begin{eqnarray}\label{eq:1solitonconstants}
q &=& \left(\begin{array}{ccc}1 + \frac{8|\alpha|^2|\delta|^2 s^2}{F} & & \frac{1}{F}[-4s\alpha \bar{\delta} (i \cA(r-m)-\cB s\cos\theta)] \\
& I_{2\times 2} & \\ 
\frac{1}{F}[4s\bar{\alpha}\delta (i \cA(r-m) + \cB s\cos\theta)] && 1+\frac{8|\alpha|^2|\delta|^2 s^2}{F} \end{array} \right),
\nonumber \\
&& \nonumber \\
\cA&=& |\alpha|^2 - |\delta|^2, \quad  \cB = |\alpha|^2 + |\delta|^2, \quad F= \cA^2((r-m)^2+s^2)-\cB^2s^2\sin^2\theta.
\end{eqnarray}
}
\end{enumerate}

At this point, we would like to systematically deduce whether particular choices of constants $\alpha, \delta$ lead to recognizable solutions, such as Kerr spacetimes.  We use this formulation of the harmonic map to develop our control theory approach.

\subsection{Control Theory meets GR: an asymptotic calculation}

In the solution-generating algorithm, it is necessary to invert the matrix $A$ to solve the linear system $AU^* = B^*$ appearing in \eqref{eq:aub} and \eqref{eq:inversion} for the unknown vectors $\bu_k$.  However, for each soliton added into the system, that is, for each $\varpi$, the matrices $U, A$, and $B$ grow in considerably size (order $N^2$).  The plan will be to replace $(r, \theta)$ with coordinates at infinity $(R, \Theta)$ and write down an asymptotic expansion for $A$ in powers of $\frac{1}{R}$, for which truncations of the matrix will be more readily inverted.  This is made possible by that fact that since the roots $\la_1, \la_{2}$ can be expressed in terms of the elliptical coordinates 
\beq
\la_1 =   (r+m +i s)(\cos\theta - 1),\quad \la_{2} =   (r+m - i s) (\cos\theta +1),
\eeq
we have $|\la_j| \to \infty$ as $|\bx| \to \infty$ (since $r_j \to \infty$ as well) for all $j$.

First observe that in the coordinates $(R, \Theta)$, the matrix $A$ can be expressed as $A = D + \frac{1}{R}N$, where $D$ is a diagonal matrix and $N$ is a nilpotent skew-hermitian matrix, $N^* = -N$.  Consequently,
\begin{eqnarray}
A&=& D+ \frac{1}{R}N = D(I + \frac{1}{R} D^{-1}N) \\
\Rightarrow  A^{-1} &=& (I + \frac{1}{R}D^{-1}N)^{-1}D^{-1} = D^{-1} -\frac{1}{R} D^{-1}ND^{-1} + \frac{1}{R^2}D^{-1}ND^{-1}ND^{-1} - \cdots \nonumber
\end{eqnarray}
This, in turn, allows us to deduce constraint equations on the constants $\alpha, \delta$ by explicitly inverting $A$ in \eqref{eq:vk} using truncations of this expansion.  In particular
\beq
U = -B D^{-1} + \frac{1}{R}B D^{-1} ND^{-1} -\frac{1}{R^2} BD^{-1} (ND^{-1})^2 + \frac{1}{R^3}BD^{-1} (ND^{-1})^3 - \cdots
\eeq
can be calculated to various desired orders of $R$.  

In our example, we shall truncate the expansion of $U$ at the $\frac{1}{R^2}$ term and, recalling the definitions of $\cA, \cB$ from \eqref{eq:1solitonconstants}, we use
\beq
B=\left[ \begin{array}{cc} -\alpha & -\alpha \\ \delta & -\delta \end{array} \right], \,\, D = \frac{\cA}{2is} \left[ \begin{array}{cc} \frac{1}{\cos \theta -1} & 0 \\ 0 & \frac{-1}{\cos \theta +1} \end{array} \right], \,\, \frac{1}{R}N= \frac{\cB}{2}\left[ \begin{array}{cc} 0 & \frac{-1}{R-m+is} \\ \frac{1}{R-m-is} & 0 \end{array} \right].
\eeq

Calculating $U$ explicitly to obtain the column vectors $\bu_j$ and combining them with the constant row vectors $\bv_j$, we compute each component $\frac{1}{\la_j}\bu_j\bv_j^*$ of the new (truncated) mapping $q^{(R)}$.  Note that each $\bu_j\bv_j^*$ contains a prefactor of $\frac{1}{\la_j} \sim \frac{1}{R}$, resulting in the following components for
$q^{(R)} \stackrel{\cdot}{=} I_{2\times 2} - \frac{1}{\la_1}\bu_1 \bv_1^* - \frac{1}{\la_2}\bu_2 \bv_2^*$ in \eqref{eq:1solitonsolq}, or equivalently \eqref{eq:1solitonconstants}:
\begin{eqnarray}
q_{11}^{(R)} &=& 1+ \frac{8s^2 |\alpha|^2 |\delta|^2}{\cA^2 [(R-m)^2+s^2]} -\frac{4\cB^2s^4 \sin^2 \theta |\alpha|^2}{\cA^3[(R-m)^2+s^2]^2} \nonumber \\
q_{12}^{(R)} &=& \frac{4s\alpha \overline{\delta}}{\cA^3[(R-m)^2+s^2]} \left\{ \cA\cB s\cos\theta [(R-m)^2 +s^2] \right. \nonumber \\
&& \qquad\qquad
\left. -i(R-m) \left( \cA^2[(R-m)^2+s^2]+\cB^2s^2\sin^2\theta \right)  \right\} \nonumber \\
 q_{21}^{(R)} &=& \frac{4s\overline{\alpha} \delta}{\cA^3[(R-m)^2+s^2]} \left\{ \cA\cB s\cos\theta [(R-m)^2 +s^2] \right.  \\
&& \qquad\qquad
\left. +i(R-m) \left( \cA^2[(R-m)^2+s^2]+\cB^2s^2\sin^2\theta \right)  \right\} \nonumber \\
 q_{22}^{(R)} &=& 1+ \frac{8s^2 |\alpha|^2 |\delta|^2}{\cA^2 [(R-m)^2+s^2]} +\frac{4\cB^2s^4 \sin^2 \theta |\delta|^2}{\cA^3[(R-m)^2+s^2]^2}. \nonumber 
\end{eqnarray}
We apply the Cayley Transform $q^\prime = QqQ^*$, with $Q=\frac{1}{\sqrt{2}}\left[ \begin{array}{cc} 1 & i \\ i & 1 \end{array} \right]$, to adjust from the target space $G/K=SU(1,1)/S(U(1) \times U(1))$ to the real hyperbolic plane  $\mathbb{H}_\mathbb{R} = SL(2, \RR)/SO(2)$, in order to compare the resulting mapping with that of the Ernst formulation.  Recall that in this setting, \eqref{eq:hyperbolic-metric} and \eqref{eq:Xdef} dictate that $X = \frac{1}{q_{22}^\prime}$, so we need only expand the $q_{22}^\prime$ entry in $R$ in order to devise constraints on the constants $\alpha, \delta$.  For the truncated mapping $q_{22}^{(R)}$,
\begin{equation}
X^{(R)} \sim 1+ \frac{-4sn_1}{\cA R} + \left( \frac{-4sn_1m}{\cA} + \frac{16sn_1^2-8s|\alpha|^2|\delta|^2 +4s^2 n_2\cB \cos \Theta}{\cA^2}  \right) \frac{1}{R^2},
\end{equation}
where we have set $\alpha \overline{\delta} = n_1 + in_2$.  By the definitions of $\cA, \cB$ in \eqref{eq:1solitonconstants}, this implies $\cB^2 - \cA^2 = 4(n_1^2+n_2^2)$.

The specification of the ADM mass as $M$, and the condition of vanishing dipole moment for the metric in the coordinate system $(R,\Theta)$ impose natural restrictions on the coefficients of the expansion in $\frac{1}{R}$ and $\frac{1}{R^2}$, namely
\begin{equation}\label{eq:firstADMdipole}
\frac{-4sn_1}{\cA} = -2M, \quad   \frac{-4sn_1m}{\cA} + \frac{16sn_1^2-8s|\alpha|^2|\delta|^2 +4s^2 n_2 \cB\cos \Theta}{\cA^2} = 0.
\end{equation}
Subjecting the second equation in \eqref{eq:firstADMdipole} to the first reduces the conditions to
\begin{equation}
\frac{-4sn_1}{\cA} = -2M, \quad   \frac{4s^2 n_2 (\cB \cos \Theta -2 n_2)}{\cA^2} = 2M^2-2Mm.
\end{equation}
Here, $a, m, s$ are free coordinate parameters, with $m, s > 0$.   Now, since these equations must hold for all $\Theta$, we have
\beq
 \frac{4s^2 n_2 \cB \cos \Theta}{\cA^2}= 0 , \quad \frac{8s^2 n_2^2}{\cA^2}= 2M^2 - 2Mm,
\eeq
whereby $n_2=0$ and thus $M=m$.  Returning to the relation between $\cA, \cB$ and $n_1, n_2$, it is easy to verify that 
\beq
\cB^2-\cA^2 = 4 n_1^2 = 4 \left(\frac{2m\cA}{4s}\right)^2 \quad \Rightarrow \quad \cB^2=\cA^2\frac{m^2 + s^2}{s^2}, 
\eeq
yielding a real scaling invariance in the choice of complex constants $\alpha$ and $\delta$, via  (see Remark \ref{rem:scaleinvariance}).  Making the choice $\cA = s$ gives $\cB = a := \sqrt{m^2+s^2}$,  and we have automatically that $n_1 = \pm \frac{ m}{2}$ (and $n_2 = 0$ is imposed already), recovering the (naked) Kerr family, as
\beq
X= \frac{r^2-2mr+a^2\cos ^2 \theta}{r^2 + a^2 \cos ^2 \theta}, \quad Y = \frac{2ma \cos \theta}{r^2 + a^2 \cos \theta}, \quad a = \sqrt{m^2 +s^2}.
\eeq

We note that the quantity $F$ appearing in \refeq{eq:1solitonconstants} is the determinant of the linear system \refeq{eq:inversion}, and that with the above choices for $\cA$ and $\cB$, it vanishes precisely when $r=0$ and $\theta = \pi/2$, i.e. on the ring singularity of the Kerr metric.  In inverting the coefficient matrix of \refeq{eq:inversion} we have assumed $F\ne 0$, which is assured for $r$ large enough, provided $\cA\ne 0$ (i.e. the extremal Kerr is also excluded from our setup.) On the other hand, the analyticity in $(r,\theta)$ of the resulting expressions for $(X,Y)$ ensures that by taking the analytic continuation of the solution one can recover the maximal  extension of the (naked) Kerr spacetime we have hereby obtained.

\subsection{Einstein-Maxwell equations}
Since the Einstein-Maxwell equations also have a harmonic maps formulation, as in Example \ref{eg:EMM}, the vesture method can equally well be used to construct new solutions, comparing with \cite{Ale80,EGK84}.

For the $N=1$ case, dressing the Minkowski seed $q_0(\bx)=I_{3 \times 3}$ with one soliton, the roots of $\la_{1,2}$ of $p(\bx, \la)$ are exactly as in the previous example.  Thus, set $\bv_1=[\alpha, \beta, \gamma]^t$ for $\alpha, \beta, \gamma \in \CC$, so that $\bv_2=\Ga \bv_1=[\alpha, \beta, -\gamma]^t$, for $\Ga= \diag(1,1,-1)$.  The resulting linear system in \eqref{eq:aub} becomes
\beq\label{linsys2}
\left[ \begin{array}{cc} \frac{\cA}{\la_1 - \overline{\la_1}} &  \frac{\cB}{\la_1 - \overline{\la_2}} \\  \frac{\cB}{\la_2 - \overline{\la_1}} &  \frac{\cA}{\la_2 - \overline{\la_2}} \end{array} \right] \left[ \begin{array}{c} \bu_1^*  \\ \bu_2^*  \end{array} \right] = \left[ \begin{array}{ccc} -\overline{\alpha} & -\overline{\beta} & \overline{\gamma} \\ -\overline{\alpha} & -\overline{\beta} & -\overline{\gamma} \end{array} \right],
\eeq
where $\cA := |\alpha|^2 +|\beta|^2 - |\gamma|^2$, and $\cB :=  |\alpha|^2 +|\beta|^2+ |\gamma|^2$.  Isolating each of the vectors $\bu_1, \bu_2$, the harmonic map is given by 
\begin{eqnarray}\label{fnlsol21} 
q(\bx) &=& I_{3\times 3} - \frac{1}{\la_1}\bu_1 \bv_1^* - \frac{1}{\la_2} \bu_2 \bv_2^* \\
	  &=&  \left(\begin{array}{ccc}
1 + \frac{8|\alpha|^2|\gamma|^2 s^2}{F} & \frac{1}{F}[8\alpha\bar{\beta}|\gamma|^2s^2] & \frac{1}{F}[-4s\al \bar{\gamma} (i \cA(r-m)-\cB s\cos\theta)] \\
q_{2,1} & 1 + \frac{8|\beta|^2|\gamma|^2 s^2}{F} & \frac{1}{F}[-4s\beta\bar{\gamma} (i \cA(r-m) - \cB s\cos\theta)] \\
q_{3,1} & q_{3,2}  & 1+\frac{8(|\alpha|^2+|\beta|^2)|\gamma|^2 s^2}{F}
 \end{array} \right) \nonumber
\end{eqnarray}
Here, $F:= \cA^2((r-m)^2+s^2)-\cB^2s^2\sin^2\theta$ and $q_{ji} = -\overline{q_{ij}}$.

Just as before, it is possible to transform to Boyer-Lindquist coordinates and consider an expansion of the matrix $A$ in powers of $\frac{1}{R}$.  We will spare the reader these details, since an identical application of the technique used in the Einstein vacuum case applies.  The above six-parameter family (three (complex)-parameter family $\alpha, \delta, \gamma$) of harmonic maps into $\cG_{2,1}$ contains as a special case, the three-parameter family of Kerr-Newman metrics in Boyer-Lindquist coordinates $(r,\theta)$.  To that end, we set $\cA = s$, whereby $\cB= -a:= \sqrt{m^2 + s^2 - e^2}$, and obtain a system of equations for the constants $\alpha\bar{\gamma} = n_1 + i n_2$ and $\beta\bar{\gamma}= n_3+in_4$ in which one finds $n_2=n_4=0$.  Matching the notation used in the previous example, we choose $n_1 =\frac{m}{2}$ and $n_3 = - \frac{e}{2}$ to then obtain exactly the Ernst potentials for the Kerr-Newman spacetimes (see Eq. (21.26), p.~326 in \cite{Stephani}),
\beq
\Phi = \frac{e}{r-ia \cos \theta} \qquad \varepsilon = 1- \frac{2m}{r-ia \cos \theta},
\eeq
for real parameters $e, a, m$. 

\section{Conclusions and Outlook}

In this section we discuss concrete ways to interpret, generalize and extend the integrability results as well as the solution-generating mechanism described to other physical theories of interest.

As we have seen, in each of the Einstein Vacuum and Einstein-Maxwell cases, the Kerr and Kerr-Newman families, respectively, can be recovered by dressing a Minkowski initial seed metric with one soliton.  This program has been completed for the broader class of complex Grassmann manifolds $\cG_{p,q}$.  The groups $SU(p,q)$ appear as universal covers of the conformal groups $SO(p,q)$, and the $\cG_{p,q}$-nonlinear sigma model described in \cite{SB-STZ} has found applications in other settings (e.g., \cite{NSanch82, Breitenlohner, Pomeransky}), and in particular in the study of higher-dimensional gravity.

Current work-in-progress includes calculating 2-soliton solutions to the Einstein Vacuum and Einstein Maxwell Equations.  Such calculations are not straightforward, since the resulting solution and hence its singular structure is unknown.  For instance, in the 1-soliton cases, an asymptotically flat spacetime with one singular ring must presumably belong to the Kerr family of spacetimes, by virtue of uniqueness results\footnote{Note however, that all of the known uniqueness results for the Kerr metric are in the black hole regime, and cover only the outside of the event horizon (cf. \cite{CLH12} and references therein.)  The same holds for ``double-Kerr" non-existence results (e.g. \cite{HenNeu09}, \cite{CENS11}).  }.  On the other hand, in dressing Minkowski with two poles, the location of the second pole gives rise to additional free parameters, and it is no longer clear whether any additional singularities (other than a second ring singularity) have been introduced into the spacetime, such as for example a singular ``strut" on the axis of symmetry between the centers of the two rings. 

We observe that 4-dimensional black holes arising from Kaluza-Klein Theories naturally set the stage for work in integrability of harmonic maps.  In \cite{Breitenlohner}, several supergravity (SUGRA) models are formulated precisely as harmonic maps into symmetric spaces.  Table \ref{table:Gibbons}, therefore comprises a collection of other potential gravitational theories which may be better understood by way of the integrability apparatus established here and in \cite{SB-STZ}; furthermore, calculation of new solutions using the dressing apparatus may now be within reach.  Entries marked with $\star$ are currently under investigation and results will be reported in a future work.

Notation for the nonlinear $\sigma$-models $f:\cM \rightarrow G/K$ refer to harmonic maps where $D$=dim $\cM$, $k$ denotes the number of Killing fields, and $m$ denotes the ratio of the number of supersymmetries of the minimal model;  supergravity (SUGRA) models closely related to the discussed harmonic maps are indicated in parentheses\footnote{The first two columns of Table \ref{table:Gibbons} are excerpted directly from \cite{Breitenlohner}, except the fifth row (see \cite{Virmani}).}.

\begin{table}[h]
\caption{Nonlinear $\sigma$-Models/Harmonic Maps $f: \cM \longrightarrow G/K$}
\centering
\begin{tabular}{ p{5cm}  p{3cm} p{3cm}  }
\hline\hline
Gravitational Theory & $G/K$ & $n$-Soliton Dressing \\[1ex]
\hline\hline
& & \\[-2.5ex]
Einstein vacuum & &   \\[-1ex]
$D=3+1$, $k=2$ & $\frac{SU(1,1)}{S(U(1)\times U(1))}$ & Kerr family  \\[-0.5ex]
($m=2$ SUGRA) & &  \\[1ex]
Einstein-Maxwell & & \\[-1ex]
$D=3+1$, $k=2$  & $\frac{SU(2,1)}{S(U(2)\times U(1))}$ & Kerr-Newman  \\[-0.5ex]
($m=2$ SUGRA) & &  \\[1ex]
Einstein gravity & $\frac{SL(n+2)}{SO(n+2)}$ & $\frac{SU(p,q)}{S(U(p)\times U(q))}$,   \\[-0.3ex]
$D=n+4$, $k=n$ & &  ($n=p+q$ \cite{SB-STZ}) \\[1ex]
$D=3+1$, $k=2$ & $\frac{SO(8,2)}{SO(8) \times SO(2)}$ &  $\quad\quad\star$ \\[-1ex]
($m=4$ SUGRA) & &  \\[1ex]
$D=5$, $k=2$ or 3 & $\frac{G_{2(2)}}{SL(2,\RR)\times SL(2,\RR)}$ & Myers-Perry, \\[-1ex]
($m=1$ SUGRA) & & Cveti\v c-Youm \\[-.3ex]
&  &  (\cite{Virmani} partial)\\[1ex]
$D=3+1$, $k=2$ & $E_{8(+8)}/SO^*(16)$ &$E_{7(+7)}/SU(8)$  \\[-0.3ex]
 (m=8 SUGRA) & (\cite{Breitenlohner} partial) & $\quad\quad\star$ \\[1ex]
\hline
\end{tabular}
\label{table:Gibbons}
\end{table}
Although integrability of the two classical equations, Einstein vacuum and Einstein-Maxwell equations, has been studied (e.g., \cite{ZB-I,ZB-II, Ale80, Ale81, EGK84}), explicit construction of solutions from this unified approach does not yet  appear in the literature.  In the case of minimal five-dimensional supergravity, it would be a significant milestone to recover the Myers-Perry or Cveti\v c-Youm solutions \cite{Virmani} systematically from this technique, identifying the seed metric as well as the pole structure imposed.  More generally, explicating black-hole solutions in $d$-dimensional vacuum gravity for $d>4$ has been of recent interest both in (minimal) supergravity and in string theory \cite{EmparanReall, Virmani}.  In this context, stationary solutions possessing $d-3$ rotational Killing fields have been extensively studied; imposing an additional timelike Killing field results in effectively two-dimensional theories, to which integrability techniques may apply.  This and other applications will be pursued elsewhere.

\section*{Acknowledgments}

SB gratefully acknowledges support from M. Kiessling and the NSF through grant DMS-0807705, during which the first part of this project was conceived; this material is also based upon work supported by the NSF under Grant \#0932078000, while SB was in residence at the Mathematical Science Research Institute in Berkeley, California, during the 2013 Autumn semester.  STZ thanks the Institute for Advanced Study for their hospitality and the stimulating environment provided during Spring 2011 while the authors were working on this project.

\nocite{*}
\bibliographystyle{plain}
\bibliography{SB_STZ_05Dec}

\end{document}